\documentclass[twocolumn]{aastex631}

\usepackage{amsmath}
\usepackage{longtable}
\usepackage{booktabs}
\usepackage{hhline}
\usepackage{multirow}
\usepackage{graphicx}
\usepackage{easyReview}

\begin{document}

\title{Near-Infrared Ca {\footnotesize{II}} Triplet As An Stellar Activity Indicator: Library and Comparative Study}

\author{X.Huang}
\author{Yuji He}
\author{Bai Zhong-Rui}
\author{Hai-Long Yuan}
\author{Mingkuan Yang}
\affiliation{CAS Key Laboratory of Optical Astronomy, National Astronomical Observatories, Chinese Academy of Sciences, Beijing 100101, China}
\affiliation{School of Astronomy and Space Science, University of Chinese Academy of Sciences, Beijing 100049, China}

\author{Ming Zhou}
\author{Yiqiao dong}

\affiliation{CAS Key Laboratory of Optical Astronomy, National Astronomical Observatories, Chinese Academy of Sciences, Beijing 100101, China}

\author{Mengxin Wang}
\affiliation{CAS Key Laboratory of Optical Astronomy, National Astronomical Observatories, Chinese Academy of Sciences, Beijing 100101, China}
\affiliation{School of Astronomy and Space Science, University of Chinese Academy of Sciences, Beijing 100049, China}

\author{Han He*}
\affiliation{National Astronomical Observatories, Chinese Academy of Sciences, Beijing 100101, China}
\affiliation{School of Astronomy and Space Science, University of Chinese Academy of Sciences, Beijing 100049, China}

\author{Jinghua Zhang}
\affiliation{South-Western Institute for Astronomy Research, Yunnan University, Chenggong District, Kunming 650500, China}

\author{Yaoquan Chu}
\affiliation{University of Science and Technology of China, Hefei 230026, China}

\author{Yongheng Zhao}
\author{Yong Zhang}
\author{H.T.Zhang*}
\affiliation{CAS Key Laboratory of Optical Astronomy, National Astronomical Observatories, Chinese Academy of Sciences, Beijing 100101, China}
\affiliation{School of Astronomy and Space Science, University of Chinese Academy of Sciences, Beijing 100049, China}

\begin{abstract}
    We have established and released a new stellar index library of the Ca {\footnotesize{II}} Triplet, which serves as an indicator for characterizing the chromospheric activity of stars. The library is based on data from the Large Sky Area Multi-Object Fiber Spectroscopic Telescope (LAMOST) Low-Resolution Spectroscopic Survey (LRS) Data Release 9 (DR9). To better reflect the chromospheric activity of stars, we have defined new indices $R$ and $R^{+}$. The library includes measurements of $R$ and $R^{+}$ for each Ca {\footnotesize{II}} infrared triplet (IRT) from 699,348 spectra of 562,863 F, G and K-type solar-like stars with Signal-to-Noise Ratio (SNR) higher than 100, as well as the stellar atmospheric parameters and basic information inherited from the LAMOST LRS Catalog. We compared the differences between the 3 individual index of the Ca {\footnotesize{II}} Triplet and also conducted a comparative analysis of $R^{+}_{\lambda8542}$ to the Ca {\footnotesize{II}} H\&K S and $R^+_{HK}$ index database. We find the fraction of low active stars decreases with $T_{eff}$ and the fraction of high active first decrease with decreasing temperature and turn to increase with decreasing temperature at 5800K. We also find a significant fraction of stars that show high activity index in both Ca {\footnotesize{II}} H\&K and IRT are binaries with low activity,
     some of them could be discriminated in Ca {\footnotesize{II}} H\&K $S$ index and $R^{+}_{\lambda8542}$ space.  
    This newly stellar library serves as a valuable resource for studying chromospheric activity in stars and can be used to improve our comprehension of stellar magnetic activity and other astrophysical phenomena.
\end{abstract}

\section {Introduction}

Stars with outer convective envelopes tend to exhibit magnetic activity. Star spots and faculae in the photosphere, plages in the chromosphere, X rays in the corona are all related to magnetic activity.
Studies of stellar activity are essential for improving our understanding of stellar dynamo models and the related studies such as the stellar age and rotation or activity relation, stellar flare and stellar activity cycle. On the other hand, stellar activity is important for exoplanets studies, since magnetic activity especially flares will have an impact on planetary habitability \citep{shields2016habitability, howard2018first, lillo2022kobe}. Also, jitters in both photometry and radial velocity measurement caused by stellar magnetic activity will hinder the detection of earth like exoplanet \citep{wright2005radial}. Finding stars with low activity
is crucial to those low mass exoplanets detecting.

The emission core of lines originated from the chromosphere can serve as indicators to quantify the activity.
One well-known measure of activity is the Ca {\footnotesize{II}} H\&K $S_{MWO}$ index, proposed by the Mount-Wilson Observatory \citep{wilson1968flux}. However, the photosphere also contributes to the Ca {\footnotesize{II}} H\&K lines flux, and the contribution varies with effective temperatures, leading to potential misestimation of the stellar activity. To overcome this issue, \citet{linsky1979stellar} proposed the $R'_{HK}$ index, which subtracts the empiprical photospheric flux from the flux. Building on the $R'_{HK}$ index, \citet{mittag2013ii, mittag2019magnetic} proposed the $R^+_{HK}$ index, which subtracts the basal flux in addition to the photospheric flux. $H_\alpha$ line can also serve as an indicator of activity, and is more suitable for late-type stars than Ca {\footnotesize{II}} H\&K \citep{cincunegui2007halpha}. They defined the $S^+$ index for $H_\alpha$, which correlates well with the $S_{MWO}$ index. 

The Ca {\footnotesize{II}} IRT lines represent another set of indices of activity:
    \begin{equation*}
        8498.0\mathring{A} \ 4 \ ^{2}P_{\frac{3}{2}}-3 \ ^{2}D_{\frac{3}{2}},
    \end{equation*}
    \begin{equation*}
        8542.1\mathring{A} \ 4 \ ^{2}P_{\frac{3}{2}}-3 \ ^{2}D_{\frac{5}{2}},
    \end{equation*}
    \begin{equation*}
        8662.1\mathring{A} \ 4 \ ^{2}P_{\frac{1}{2}}-3 \ ^{2}D_{\frac{3}{2}},
    \end{equation*}
absorptions due to the Ca {\footnotesize{II}} IRT lines are clearly visible in the atmosphere of cool stars \citep[see][chap. 6]{tennyson2019astronomical}. The Ca {\footnotesize{II}} IRT emission lines core are formed in the lower chromosphere through subordinate transitions between the excited levels of Ca {\footnotesize{II}} $4 \ ^{2}P_{\frac{3}{2}, \frac{1}{2}}$ and meta-stable $3 \ ^{2}D_{\frac{3}{2}, \frac{5}{2}}$. These lines are mostly collision controlled \citep{de2021stellar}, and are highly sensitive to the ambient temperature \citep{Cauzzi2008}. They are indicator of stellar chromospheric activity, as demonstrated by \citet{linsky1979stellar}.
\citet{linsky1979stellar} proposed using Ca {\footnotesize{II}} $\lambda 8542$ as an activity indicator, while \cite{andretta2005ii} defined the $R_{IRT}$ index based on the central depression in the Ca {\footnotesize{II}} IRT lines, taking into account rotational broadening. \citet{notsu2015high} used $r_{0}(IRT)$, which is the residual flux normalized by the continuum at the line
cores of IRT lines, and $H_{\alpha}$ to study superflare and suggested that the brightness variation of superflare stars can be explained by the rotation with large starspots. \citet{vzerjal2013chromospherically} use observed spectra of non-active stars as template, and
measure the template subtracted equivalent width(EW) of the Ca {\footnotesize{II}} IRT lines to represent the stellar activity.

It is important to built large databases to statistically understanding the physical mechanisms of stellar magnetic activity. As a series of work, we have already built large sample databases of stellar activity of solar like stars using Ca {\footnotesize{II}} H\&K \citep{zhang2022stellar} and $\rm H_{\alpha}$ \citep{he2023h} index based on LAMOST spectra. In the current work, we will build a stellar activity database of F, G, K stars based on the measurement of Ca {\footnotesize{II}} IRT lines.

LAMOST, the Large Sky Area Multi-Object Fiber Spectroscopic Telescope located in Xinglong, China, offers low-resolution spectra with a resolving power of $\lambda/\Delta \lambda=1800$ covering the wavelength range of 3700-9100 $\mathring{A}$ \citep{zhao2012lamost}. Additionally, it provides Mid-Resolution Spectra (MRS) with $R\sim 7500$ in 4950-5350 $\mathring{A}$, 6300-6800$\mathring{A}$ band. The observed data was first reduced by LAMOST 2D pipeline \citep{bai2017sky, bai2021first}, then LAMOST stellar parameter pipeline \citep{wu2011automatic}. The released data including extracted spectra files as well as the stellar parameters are available at the LAMOST website, \url{http://www.lamost.org}.

There have been several studies of stellar activity based in LAMOST data. For example, \citet{zhang2020magnetic} employed the $R^{+}_{HK}$ index to investigate the relationship between stellar activity, period, and the amplitude of brightness variation; \citet{he2023h} measured the $R_{H_{\alpha}}$ index using LAMOST MRS; \citet{zhang2022stellar} established Ca {\footnotesize{II}} H\&K $S$ index database base on LAMOST LRS; \citet{karoff2016observational} explored superflares using the $S$ index and found that superflare stars are characterized by enhanced activity; \citet{zhang2019stellar} proposed that stellar chromospheric activity indices can be used to roughly estimate stellar ages for dwarfs. The above studies are based on the measurement of Ca {\footnotesize{II}} H\&K or $H_{\alpha}$, the capability of Ca {\footnotesize{II}} IRT lines has not been fully explored yet.

In this study, we concentrate on Ca {\footnotesize{II}} IRT lines of solar-like stars, all the spectra utilized in our research come from the LAMOST LRS DR9 database. Due to the low spectral resolution, the line core emission is not sensitive to equivalent width (EW) and may be compromised by deviations in rotation velocity estimations. Instead, we introduce a new $R$ index that specifically considers the flux near the center of spectral lines. To remove the photospheric flux components, we employed the \texttt{BT-Settl} stellar spectral models (\citep{allard1997model, allard2011model, allard2013progress} and calculate the template subtracted index, $R^+$, to represent pure activity levels. Additionally, we compare our results with the existing database of Ca {\footnotesize{II}} H\&K lines and discuss the nature of stars in the Ca {\footnotesize{II}} H\&K and IRT activity index distribution. 

This paper is organized in five sections. Section 2 introduces the data selection criteria, while Section 3 defines the indices $R$ and $R^+$ and provides a detailed description of the data processing steps. Section 4 shows the detail of our database. In Section 5 we compared the strengths of the three lines, discusses the relationship and differences between the indices measured from Ca {\footnotesize{II}} H\&K.   Section 6 is the summary.

\section{Data Preparation}

Our analysis focuses on F, G and K-type solar-like stars, with all stellar parameters sourced from the catalog: \texttt{LAMOST LRS Stellar Parameter of A, F, G, and K Stars} (\textit{AFGK} Catalog) (\url{http://www.lamost.org/dr9/}). To be comparable with the previous Ca {\footnotesize{II}} H\&K index work of \citet{zhang2022stellar}, the following parameter restrictions are adopted:

\begin{enumerate}
\item $100 \leq SNR_i,\ SNR_z$. This is to ensure the high quality of the Ca {\footnotesize{II}} IRT lines located between i \& z band.
\item $4800K \leq T_{eff} \leq 6800K$, This criterion is same as \citet{zhang2022stellar}, the temperature range of solar-like stars covers most F, G, K samples in the \textit{AFGK} Catalog. 
\item For surface gravity, the empirical formulas of \citet{zhang2022stellar} is adopted to select main sequence stars:
    \begin{flalign*}
        &5.98 - 0.00035T_{eff} \leq \text{log}\ g \leq 5.0
    \end{flalign*}
\end{enumerate}

After rejecting spectra with issues such as fiber failure at the IRT bandpass, heavy skylight pollution, and wavelength calibration failure, we selected a total of 699,348 spectra from the LAMOST database. As there are multiple visits for the same star, these spectra are from 562,863 stars. The number of spectra cross-correlated with the previous work of Ca {\footnotesize{II}} H\&K $S$ and $R^{+}_{HK}$ index databases is listed in Table \ref{tab1}.

\begin{table}[!htbp]
    \centering
    \caption{\textbf{Ca {\footnotesize{II}} index Database Using LAMOST Data }}
    \begin{tabular}{ l c c }
        \toprule
        \textbf{\quad Database} & \textbf{Spectra Number} & \textbf{Common Spectra} \\ 
        \hline
        Ca {\footnotesize{II}} IRT $R$, $R^+$ & 699348 & - \\
        Ca {\footnotesize{II}} H\&K $S$  & 1330654 & 574780 \\
        Ca {\footnotesize{II}} H\&K $R^+_{HK}$   & 59816 & 14028 \\
        \hline
    \end{tabular}
    \tablecomments{IRT $R$, $R^+$ Index Database is given in this work. $S$ Index database is provided by \citet{zhang2022stellar} and $R^+_{HK}$ Index is provided by \citet{zhang2020magnetic}.}
    \label{tab1}
\end{table}

\section{Method}
\subsection{Index definitions}
We defined $R$, $R^+$ index for each line of Ca {\footnotesize{II}} IRT as following equations:

\begin{equation}
    R = \frac{1}{\lambda_2 - \lambda_1}\int^{\lambda_2}_{\lambda_1}\frac{F_{o}(\lambda)}{C_{o}(\lambda)} \ d\lambda,
    \label{eq1}
\end{equation}

\begin{equation}
    R^+ = \frac{1}{\lambda_2 - \lambda_1}\int^{\lambda_2}_{\lambda_1}\frac{F_{o}(\lambda)}{C_{o}(\lambda)}-\frac{F_{m}(\lambda)}{C_{m}(\lambda)} \ d\lambda,
    \label{eq2}
\end{equation}

where $F(\lambda)$ is the spectrum, $C(\lambda)$ is the linear function fitting the local continuum at IRT bandpass, and subscript o and m stand for observation and model respectively.
 $F(\lambda)/C(\lambda)$ is the normalized spectrum. $\lambda_1$, $\lambda_2$ are the starting and ending wavelength of the sampling range, which is 1$\mathring{A}$ around the central wavelength of each Ca {\footnotesize{II}} IRT lines. The corresponding central wavelengths and the sampling ranges are listed in Table \ref{tab2}. As the LAMOST spectral data points are in approximately 2$\mathring{A}$ interval, a cubic spline function is applied to interpolate the spectrum to 0.001$\mathring{A}$ interval.

\begin{table}[!htbp]
    \centering
    \caption{\textbf{Sampling Range for Ca {\footnotesize{II}} IRT Index}}
    \begin{tabular}{l c c }
        \toprule
        \textbf{Lines} & \textbf{Center$(\mathring{A})$}   & \textbf{Bandpass$(\mathring{A})$} \\ 
        \hline
        Ca {\footnotesize{II}} {$\lambda 8498$} & 8500.35 & 8549.85-8500.85 \\
        Ca {\footnotesize{II}} {$\lambda 8542$} & 8544.44 & 8543.94-8544.94 \\
        Ca {\footnotesize{II}} {$\lambda 8662$} & 8664.52 & 8664.02-8665.02 \\
        \hline
    \end{tabular}
    \tablecomments{The wavelength are in vacuum, as provided by LAMOST data release(\href{http://www.lamost.org/dr9/v1.0/search}{LAMOST LRS DR9}).\label{tab2}}
\end{table}

\begin{figure}[!htbp]
    \centering
    \includegraphics[scale=0.54]{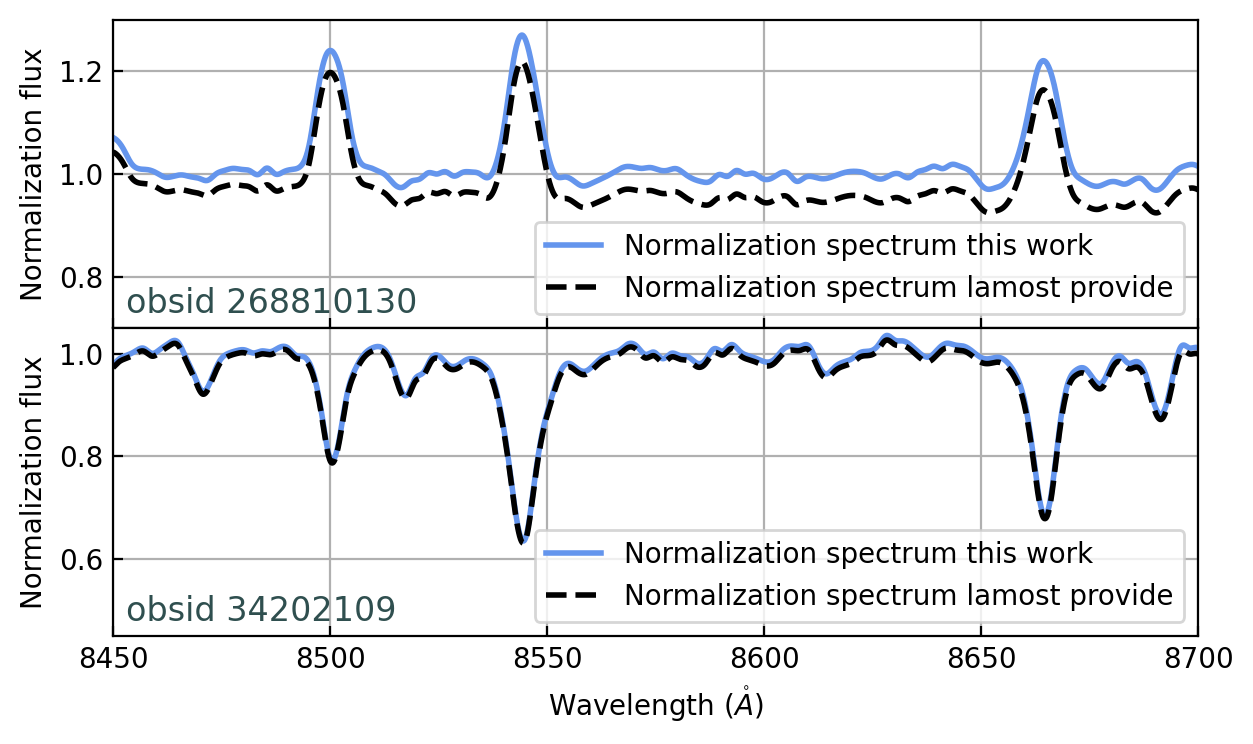}
    \caption{Comparison of different normalization in the Ca {\footnotesize{II}} IRT region. the upper panel is the emission lines spectrum and the lower is the absorption lines spectrum. The blue curve is the local normalized spectra by this work and the black curve is the global normalized spectra provided by LAMOST DR9.}
    \label{fig1}
\end{figure}

 LAMOST DR9 provides normalized spectra for most spectra, these are typically generated for the entire spectrum. To achieve better performance, we re-normalized the spectra within the IRT bandpass with a normalization method that utilizes the \texttt{LinearLSQFitter} provide by \texttt{Astropy} module, which is a linear least square fitting method \citep{robitaille2013astropy, price2018astropy, price2022astropy}. 
 Two examples are illustrated in Figure \ref{fig1} to show the difference between global and local normalization. 
 Both methods perform similarly for the absorption line spectra, but in the case of emission lines, our method clearly outperforms the LAMOST approach.

\subsection{Templates}
For late type stars. the dissipation of acoustic energy \citep{Schrijver1989} and turbulent dynamo activity from non-rotating plasma \citep{Bercik2005} in the upper photosphere contribute to the line core of Ca {\footnotesize{II}} H\&K and Ca {\footnotesize{II}} IRT lines, thus it is better to subtract this "basal" flux from the spectrum to derive the true chromosphere activity. \citet{andretta2005ii} studied the non-local thermodynamic equilibrium (NLTE) effect contribution of Ca {\footnotesize{II}} IRT lines, and found that the Central-Depression(CD) index can be affected by NLTE by more than 20\%. As our $R^{+}$ and $R$ indices are defined on narrow band of 1\AA , similar to CD index, NLTE should be consider in $R^{+}$ index to remove the basal flux. The LTE \texttt{BT-Settl} spectral model and NLTE model for Ca {\footnotesize{II}} lines \citep{allard2013progress} based on \texttt{Phoenix} \citep{husser2013new} code were applied to subtract the basal flux in Ca {\footnotesize{II}} IRT region.



The grids of \texttt{BT-Settl} templates are listed in Table \ref{tab3}. These templates were interpolated with intervals of $\Delta T_{eff} = 10K$, $\Delta\text{log}\ g = 0.01$ and $\Delta[Fe/H] = 0.01$ to ensure a precise match with our observational spectra. The templates are degraded to $R \approx 1800$ and subtracted from the observed spectra, as equation \ref{eq2}.

\begin{table}[!htbp]
    \centering
    \caption{\textbf{Parameter Space of The Grid}}
    \begin{tabular}{l c c }
        \toprule
            \textbf{Parameter} & \ \textbf{Range} & \ \textbf{Grid Size} \\ 
            \hline
            $T_{eff}$(K)   & 4800-6800  & 100 \\
            $\text{log}\ g$ & 3.5-5.0   & 0.5 \\
            $[Fe/H]$       & [-1.0,-0.5,0,0.3,0.5]  & - \\
            $[\alpha/Fe]$  & 0.0-0.4    & 0.2 \\
            \hline
    \end{tabular}
    \tablecomments{For most LAMOST spectra, $\alpha$ abundance is not offered in Dr9, the following empirical relations are used to derive $\alpha$ abundance : $[Fe/H] = 0.0, +0.3, +0.5$ with $[\alpha/Fe]=0.0$, $[Fe/H] = -0.5$ with $[\alpha/Fe]= +0.2$, $[Fe/H] = -1.0$ with $[\alpha/Fe] = +0.4$.}
    \label{tab3}
\end{table}

\begin{figure}[!htbp]
    \centering
    \includegraphics[scale=0.42]{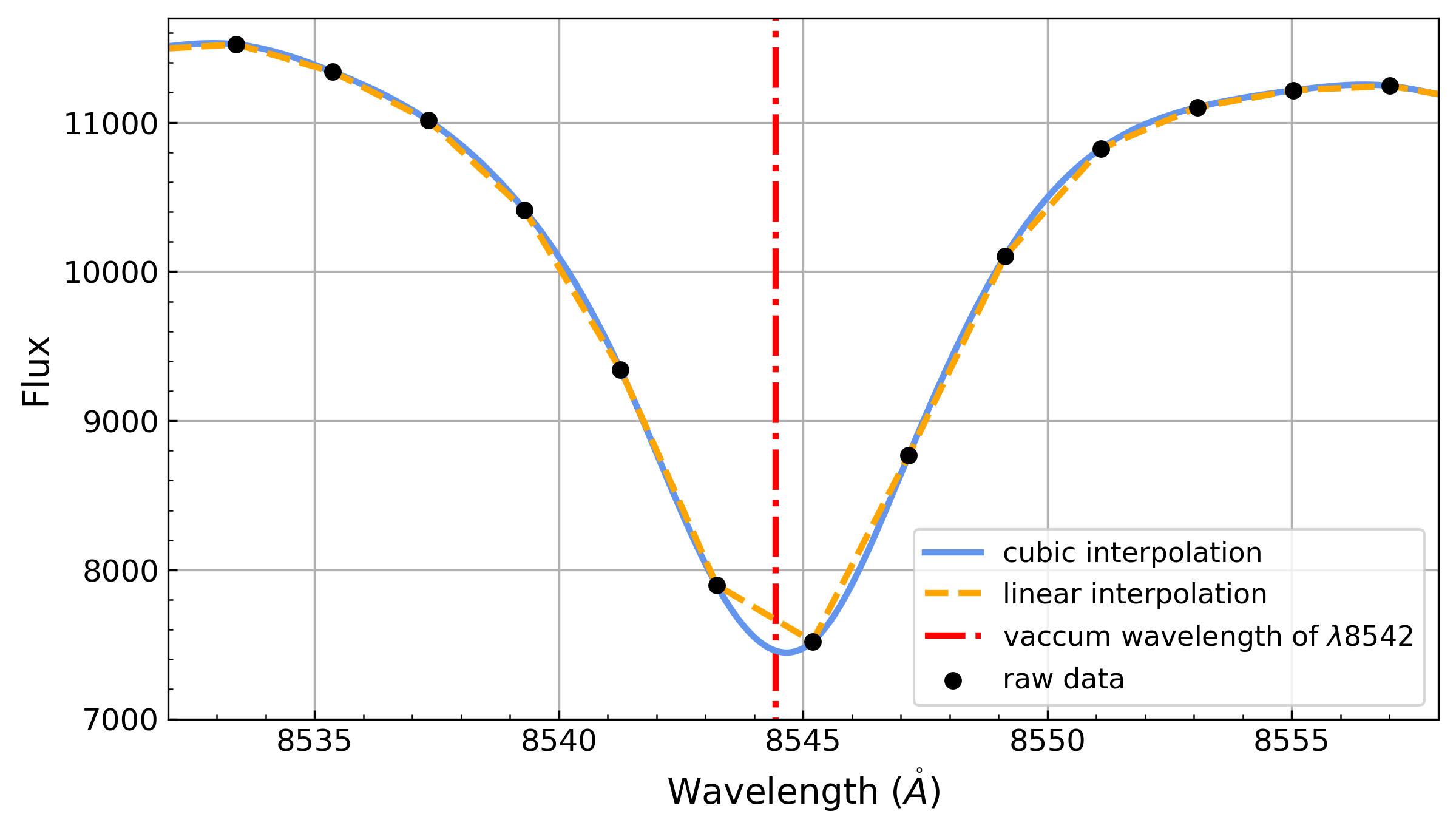}
    \caption{Difference of two interpolation method. Black dots are observed spectrum; Blue curve is the cubic spline interpolation of the spectrum; Orange dash curve is the linear interpolation; Red dot-dash line shows the vaccum wavelength of $\lambda 8542$.}
    \label{fig2}
\end{figure}

\begin{figure*}[!htbp]
    \centering
    \includegraphics[scale=0.62]{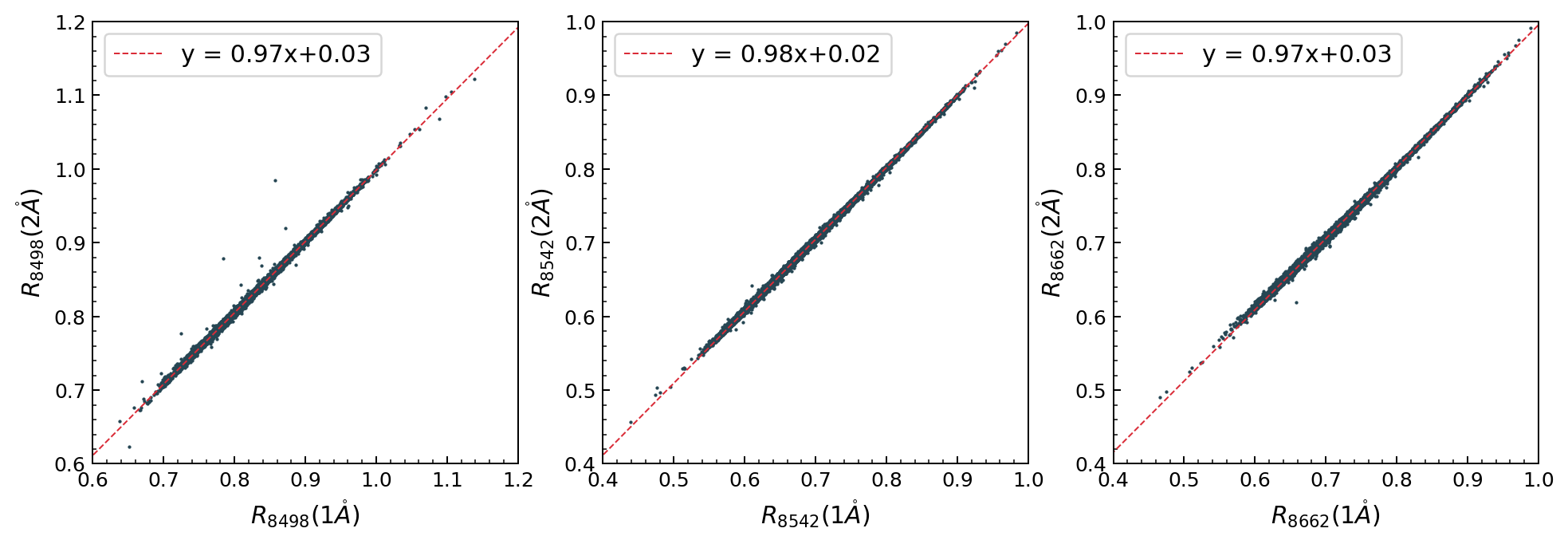}
    \caption{Comparison of $R$ index derived from 1 $\mathring{A}$ and 2 $\mathring{A}$ width respectively for each IRT line. Red dash line is obtained by least squares fitting of the data.}
    \label{fig3}
\end{figure*}

\subsection{Uncertainties Estimation}

Similar to the LAMOST Ca {\footnotesize{II}} H\&K index error budget analysis in \citet{zhang2022stellar} , for Ca {\footnotesize{II}} IRT $R$ index, we consider three factors of uncertainty as follows:

\begin{enumerate}
    
    \item Uncertainty of spectral flux. LAMOST release the targets spectrum as well as 
    the corresponding spectra of inverse variance($1/\delta^2$), which could be used to estimate the flux uncertainty: 
    
    \begin{equation}
        \delta R_{flux} = \sqrt{\frac{1}{\lambda_2 - \lambda_1}{\int^{\lambda_2}_{\lambda_1}(\frac{\delta(\lambda)}{C(\lambda)})^2} \ d\lambda},
        \label{eq3}
    \end{equation}
    where $C(\lambda)$ is the continuum, same as defined in equation \ref{eq1}.
    
    \item Uncertainty of interpolation. As the wavelength interval of LAMOST spectra is 2\AA , the spectrum are interpolated. Different interpolation method lead to the uncertainty of $R$ index, as illustrated in Figure \ref{fig2}. The uncertainty of interpolation is derived as: 
    \begin{equation}
        \delta R_{interpolation} = |R_{cubic} - R_{linear}|,
        \label{eq4}
    \end{equation}
    
    to ensure that our choice of 1 \AA\ window doesn't impact our conclusions, 
    we compared the $R$ indices of each Ca {\footnotesize{II}} IRT line measured in 1 $\mathring{A}$ window with those of the 2 $\mathring{A}$ window. For majority of targets, the difference is negligible, as shown in Figure \ref{fig3}.
    
    
    \item Uncertainty of red shift(or radial velocity). by applying $z+z_{err}$, $z$, $z-z_{err}$ offered by LAMOST DR9, we can obtain the $R_{+}$, $R$, $R_{-}$ respectively for each line, so $\delta R_{z}$ is shown as following:

    \begin{equation}
        \delta R_{z} = \frac{|R - R_{+}| + |R - R_{-}|}{2}.
        \label{eq5}
    \end{equation}
\end{enumerate}

Combining function \ref{eq3},\ref{eq4} and \ref{eq5}, the total error $\delta R$ is give by,

\begin{equation}
    \delta R = \sqrt{\delta R_{flux}^2 + \delta R_{interpolation}^2 + \delta R_{z}^2}.
    \label{eq6}
\end{equation}

\begin{figure*}[!htbp]
    \centering
    \includegraphics[scale=0.68]{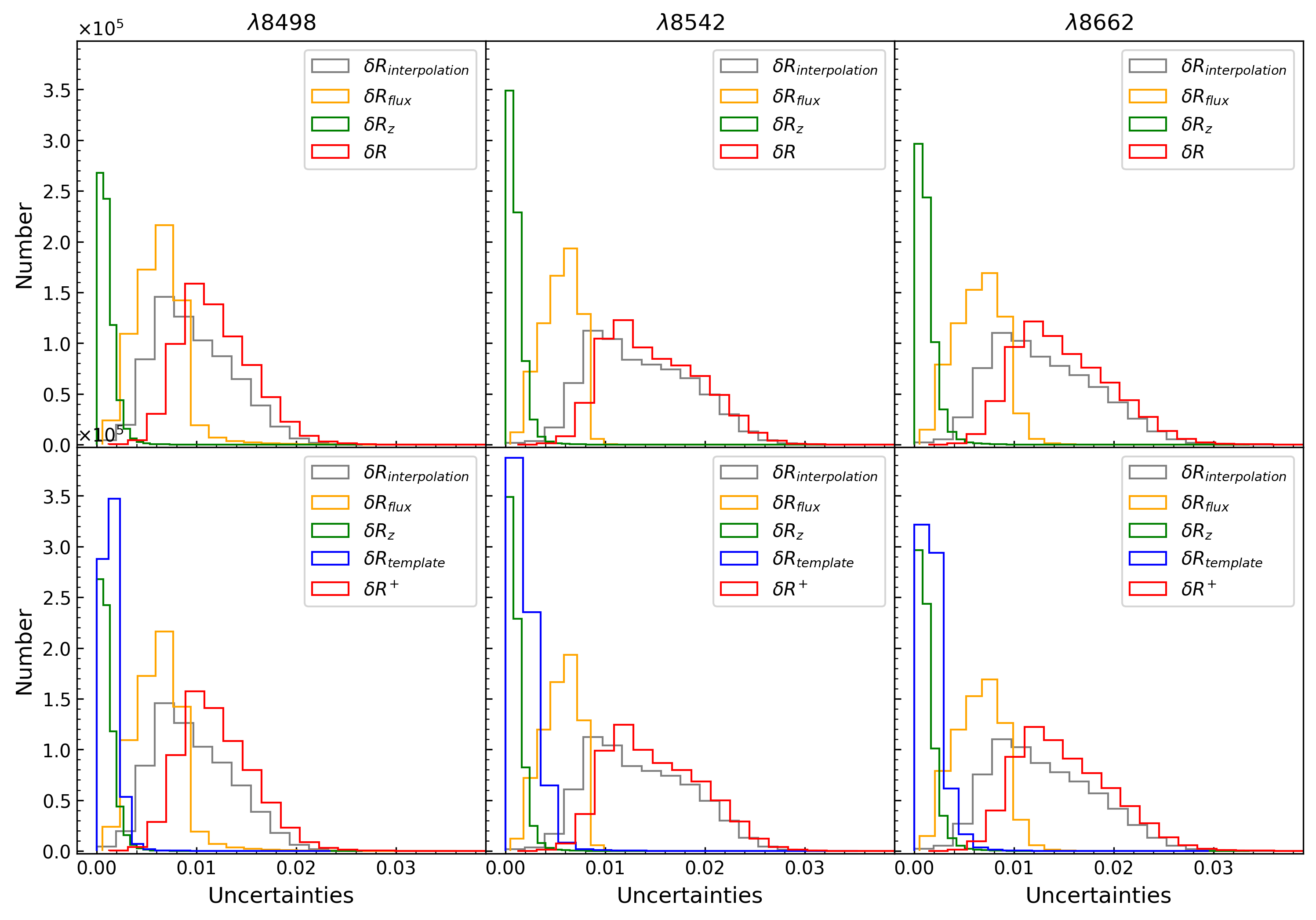}
    \caption{Distribution of uncertainties for the spectral lines $\lambda 8498$, $\lambda 8542$, and $\lambda 8662$ in three columns from left to right. Each column includes two panels, with the top one showing the distribution of uncertainty for $R$ and its individual component, and the bottom one displaying the distribution of uncertainty for $R^{+}$, both represented by the red histogram.}
    \label{fig4}
\end{figure*}

\begin{figure*}[!htbp]
    \centering
    \includegraphics[scale=0.56]{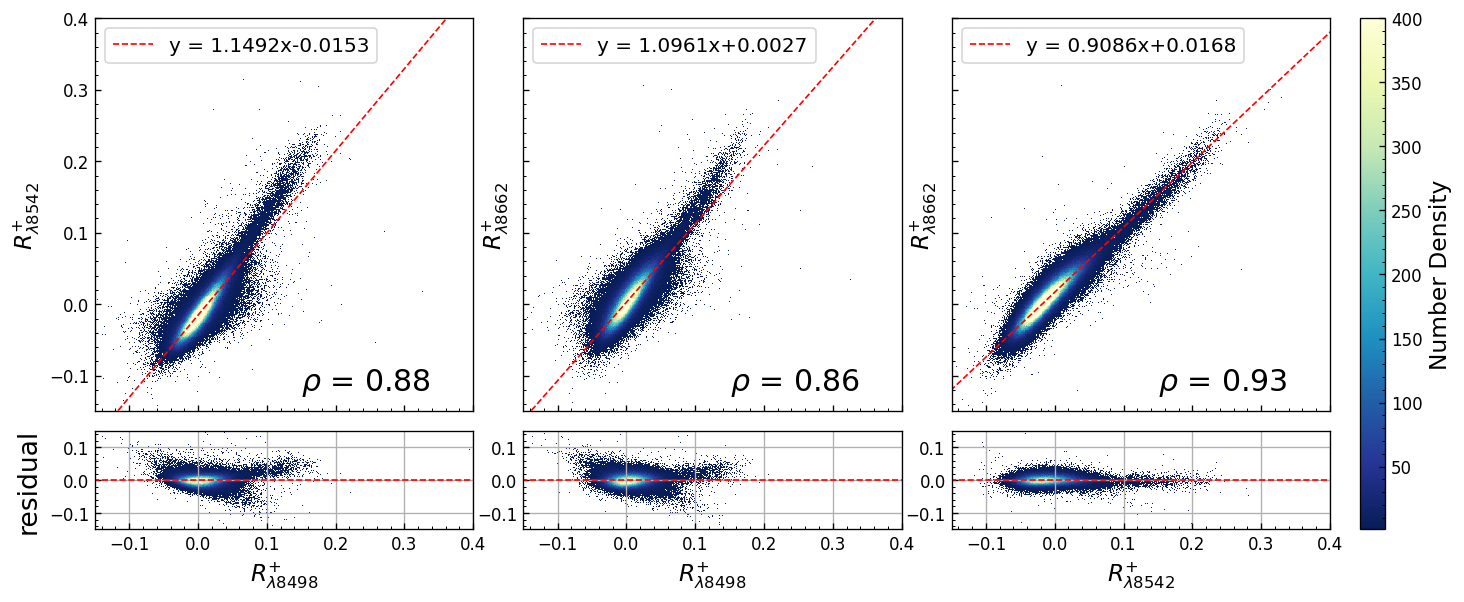}
    \caption{Linear regression was performed for each pair of $R^{+}$ values in IRT, with the corresponding residuals between the data and the fitted line shown in the lower panels. The left column displays $R^{+}_{\lambda 8498}$ - $R^{+}_{\lambda 8542}$, the middle column shows $R^{+}_{\lambda 8498}$ - $R^{+}_{\lambda 8662}$, and the right column depicts $R^{+}_{\lambda 8542}$ - $R^{+}_{\lambda 8662}$. The red dashed lines represent the regression equation obtained from fitting the data, while $\rho$ corresponds to the Pearson correlation coefficient.}
    \label{fig5}
\end{figure*}


\begin{figure*}[!htbp]
    \centering
    \includegraphics[scale=0.62]{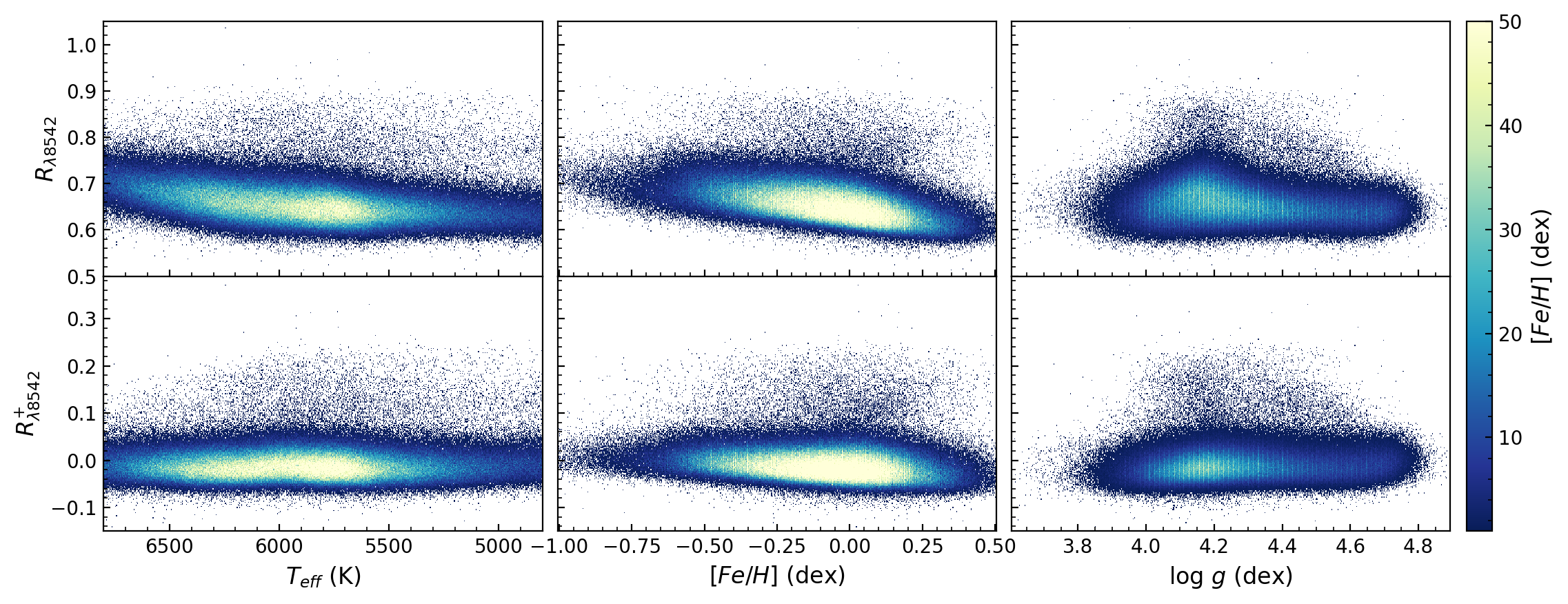}
    \caption{The distribution of $R_{\lambda 8542}$ and $R^{+}_{\lambda 8542}$ with stellar parameters. From top to bottom are $T_{eff}$, $[Fe/H]$ and $\text{log}\ g$, respectively.
    The upper section in each panel is $R$ index and the lower is for $R^{+}$, as indicated in the plot.}
    \label{fig7}
\end{figure*}

\begin{figure}[!htbp]
    \centering
    \includegraphics[scale=0.62]{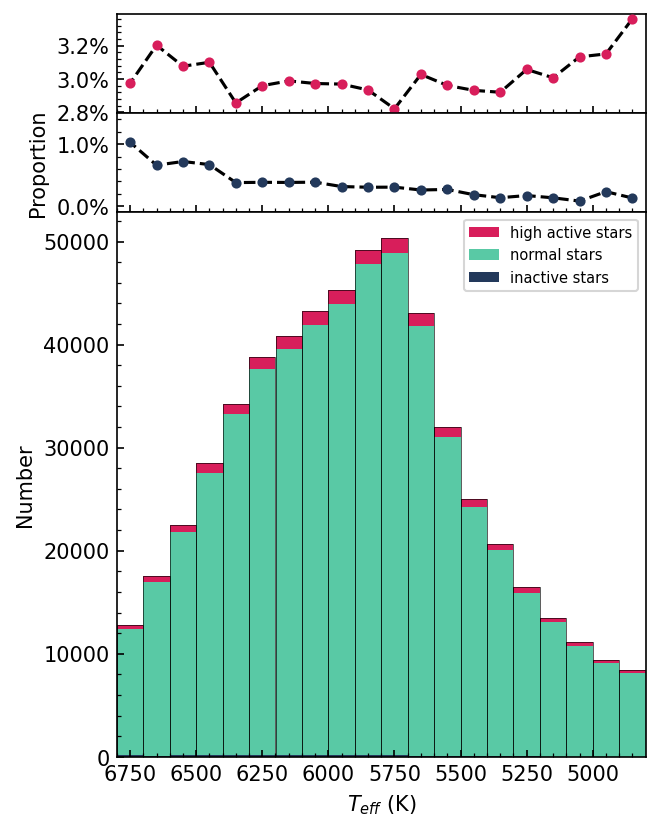}
    \caption{Top and mid panels shows the proportion of high active stars and inactive stars, respectively. The bottom panel shows the number count of different category in different temperature bins, as indicated by different color.}
    \label{fig6}
\end{figure}


For $R^+$ index, the additional uncertainty comes from the templates uncertainty. According to the stellar parameter error offered by LAMOST DR9, we calculate a serious of $R$ index for each templates around the best template, [$T_{eff}\pm \Delta T$, $\text{log}\ g\pm \Delta \text{log}\ g$, $[Fe/H]\pm \Delta[Fe/H]$]. The maximum and minimum of the template index $R_T$ are denoted as $R_{T}^{max}$ and $R_{T}^{min}$ respectively. Then uncertainty of template index $R_T$ is

\begin{equation}
    \delta R_{T} = max(|R_T - R_{T}^{max}|, \ |R_T - R_{T}^{min}|),
    \label{eq7}
\end{equation}
and the uncertainty of $R^+$ is
\begin{equation}
    \delta R^+ = \sqrt{\delta R^2 + \delta R_{T}^2}.
    \label{eq8}
\end{equation}

Figure \ref{fig4} shows the contribution of different components to $\delta R$ and $\delta R^+$, we can see that the uncertainty of $R^{+}$ is mainly dominated by the uncertainty of interpolation and flux error.

\section{Stellar Activity Database}

We calulated the $R$ and $R^+$ index and the corresponding error for 699,348 F, G and K-type spectra selected from LAMOST DR9 database. 
The results are written in a CSV form table and uploaded to the website \url{https://nadc.china-vo.org/res/r101246/}.
The description of columns of the database could be found in Table \ref{tab4}. Our R and $R^+$ index database could be used as indicator for stellar activity studies.
Theoretically, the $R^{+}$ index is close to zero for inactive stars, but there are a large fraction of stars with the $R^{+}$ index below zero (see Figure \ref{fig7}). The similar negative value are also found in GAIA \citep{2023A&A...674A..30L} and RAVE \citep{vzerjal2013chromospherically} Ca {\footnotesize{II}} IRT index measurement. We believe that the following reasons have led to this:

\begin{enumerate} 
\item The parameters of LAMOST may not have been measured accurately. 
\item Low or moderate chromospheric activity could produce some extra absorption \citep{mullan1979model, 2023A&A...674A..30L}. 
\end{enumerate}

\setlength{\tabcolsep}{0.1mm}
\begin{longtable}{l c c}
    \caption{\textbf{Columns of catalog}}\\
    \toprule
        \textbf{Column}           & \textbf{Unit}               & \textbf{Description}                                                    \\
        \hline
        obsid                     &                             & LAMOST observation identifier                                           \\
        gaia\_source\_id          &                             & Source identifier in Gaia DR3                                           \\
        gaia\_g\_mean\_mag        &                             & G mag provided by Gaia DR3                                              \\
        snri                      &                             & SNR at i band                                                           \\
        snrz                      &                             & SNR at z band                                                           \\
        ra\_obs                   & degree                      & RA of fiber point                                                       \\
        dec\_obs                  & degree                      & DEC of fiber point                                                      \\
        teff                      & K                           & $T_{eff}$, Effective temperature                                        \\
        teff\_err                 & K                           & Uncertainty of $T_{eff}$                                                \\
        logg                      & dex                         & $\text{log}\ g$, Surface gravity                                        \\
        logg\_err                 & dex                         & Uncertainty of $\text{log}\ g$                                          \\
        feh                       & dex                         & $[Fe/H]$, Metallicity                                                   \\
        feh\_err                  & dex                         & Uncertainty of $[Fe/H]$                                                 \\
        rv                        & km/s                        & $V_r$, Radial velocity                                                  \\
        rv\_err                   & km/s                        & Uncertainty of $V_r$                                                    \\
        R\_8498                   &                             & $R_{\lambda 8498}$                                                      \\
        R\_8498\_err              &                             & uncertainty of $R_{\lambda 8498}$                                       \\
        R\_8542                   &                             & $R_{\lambda 8542}$                                                      \\
        R\_8542\_err              &                             & uncertainty of $R_{\lambda 8542}$                                       \\
        R\_8662                   &                             & $R_{\lambda 8662}$                                                      \\
        R\_8662\_err              &                             & uncertainty of $R_{\lambda 8662}$                                       \\
        R\_plus\_8498             &                             & $R^{+}_{\lambda 8498}$                                                  \\
        R\_plus\_8498\_err        &                             & uncertainty of $R^{+}_{\lambda 8498}$                                   \\
        R\_plus\_8542             &                             & $R^{+}_{\lambda 8542}$                                                  \\
        R\_plus\_8542\_err        &                             & uncertainty of $R^{+}_{\lambda 8542}$                                   \\
        R\_plus\_8662             &                             & $R^{+}_{\lambda 8662}$                                                  \\
        R\_plus\_8662\_err        &                             & uncertainty of $R^{+}_{\lambda 8662}$                                   \\
        \hline
    \label{tab4}
\end{longtable}
\tablecomments{Some of the stellar parameters error or the indices error are not available in the data release, the corresponding error in the table are filled with -9999.}

\begin{figure*}[!htbp]
    \centering
    \includegraphics[scale=0.48]{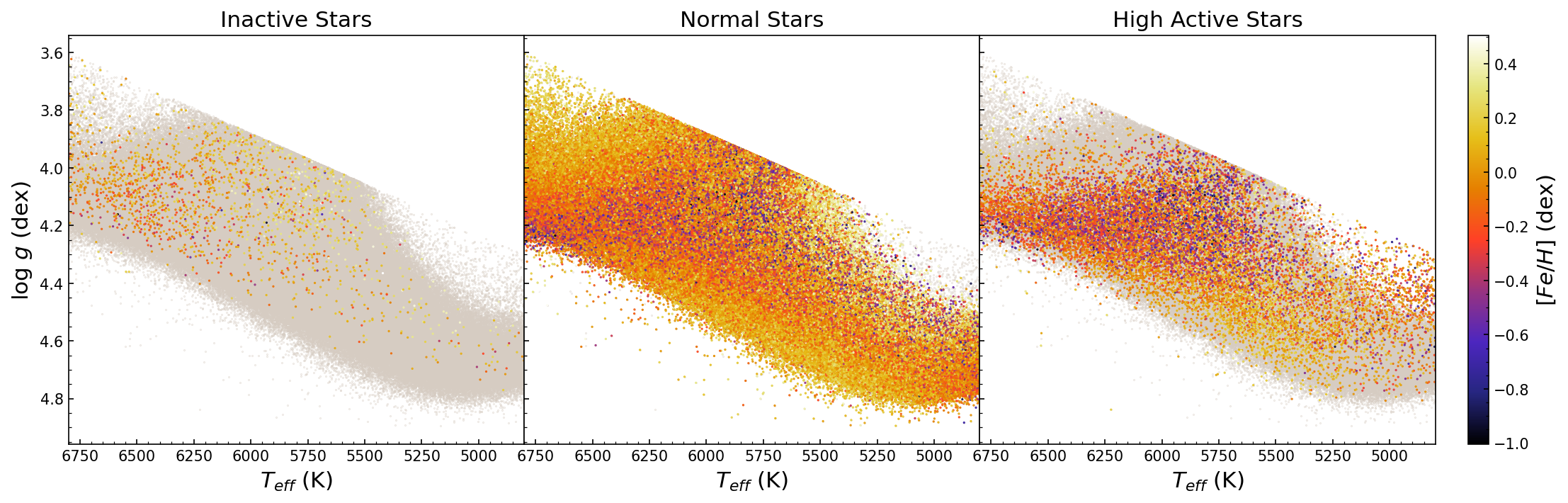}
    \caption{Distribution of inactive, normal and high active stars in the $T_{eff}$ and $Log g$ space, different color represent different [Fe/H], as indicated by the color bar.}
    \label{fig8}
\end{figure*}

\section{Discussion}

\subsection{Relationship between IRT indices and stellar parameters}

In Figure \ref{fig5}, we plot the Ca {\footnotesize{II}} IRT $R^+$ index against each other. There is a clear linear correlation in each plot. We calculated the Pearson correlation coefficient and marked at the lower part of each panel. For each pair, the ridge of the density distribution is fitted with a linear function using the \texttt{Bayesian Ridge Regression} algorithm from the \texttt{sklearn} module \citep{scikit-learn}. The functions are shown on the top of each panel of Figure \ref{fig5}. From the figure, we can see that $R^{+}_{\lambda 8542}-R^{+}_{\lambda 8662}$ exhibit the strongest linear relationship, with a higher Pearson coefficient than other
pairs. The $\lambda 8542$ line is the most opaque member of the Ca {\footnotesize{II}} IRT lines and usually considered as a better diagnostic for the chromospheric activities \citep{linsky1979stellar}.
According to the linear function slopes, the strength of $R^{+}_{\lambda 8542}$ is stronger than the other two lines, our results confirms the conclusion of \citet{linsky1979stellar} and are also consistent with the results of \citet{vzerjal2013chromospherically} and \citet{martin2017ii}.
Henceforth, we limit our discussion to $\lambda 8542$, although all the other line index are available in our database for possible use.


The distributions of $R^{+}_{\lambda 8542}$ and $R_{\lambda 8542}$ with stellar parameters are presented in Figure \ref{fig7}. Stars with low activity is also important for low mass exoplanets studies since life may more possibly exist in a planet hosted by low active star and exoplanet may be more easily discovered around low active stars than the active because both the observed lightcurve and radial velocity curve will be more stable due to less spots on the star \citep[e.g.][]{exoplanet2015,exoplanet2020}. To take a peek at the distribution of the chromospheric active and inactive stars, the star are divided into 20 temperature bins, and the number count in each bins are plotted in the bottom panel of Figure \ref{fig6}. The mean and variance of $R^{+}_{\lambda 8542}$ are calculated for each bin. Stars with $R^{+}$ index higher than 2$\sigma$ are defined as active stars and lower than 2$\sigma$ are inactive stars. The fractions of active and inactive stars are plotted in the upper 2 panels of Figure \ref{fig6}. The fraction of inactive stars decrease with temperature. While the fraction of active stars increase with the decreasing temperature below 5800K, and increase with temperature above 5800K. As there are a large fraction of high $R$ index stars are actually binaries (see Section \ref{sec:sind} below), the increasing of active star fraction with temperature may reflect the increasing binary fraction with mass rather than the increasing activity. Further work is needed to make it clear. The distributions of active, inactive and normal stars in the stellar parameter space are shown in Figure \ref{fig8}. The inactive stars show high metallicity in Figure \ref{fig8} , indicating that they are thin disk population, similarly, the low metallicity population in the active stars plot may possibly comes from the local thick disk population. As some stars were observed several times by LAMOST, for Figure \ref{fig7}, \ref{fig6} and \ref{fig8}, only one spectrum were kept for stars with multiple visits to ensure the fraction is not biased by repeat count. As the stellar activity is a complicated function of mass, age, metallicity and rotation, which is beyond the scope of the current paper, we will leave the detail analysis for the future work. 


\subsection{Comparing with S index\label{sec:sind}}
\begin{figure}[!htbp]
    \centering
    \includegraphics[scale=0.57]{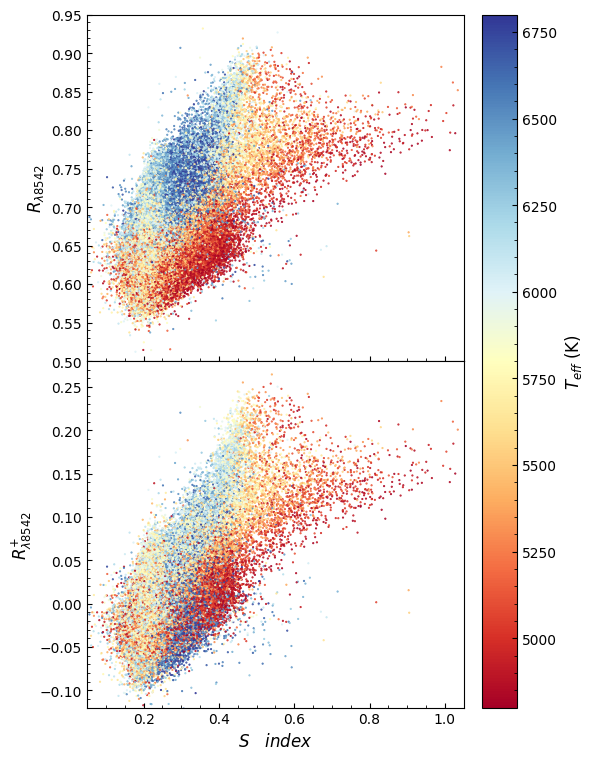}
    \caption{Top panel is the relationship of $S$ index and $R_{\lambda 8542}$, bottom panel is the relationship of $S$ index and $R^{+}_{\lambda 8542}$. Colors of each point show the temperature.}
    \label{fig9}
\end{figure}

Comparing our database with Ca {\footnotesize{II}} $H\&K$ $S$ index of \citet{zhang2022stellar}, there are 0.58 million spectra in common(Table \ref{tab1}). The $S$ index VS $R^{+ }_{\lambda 8542}$ and $S$ VS $R_{\lambda 8542}$ are plotted in Figure \ref{fig9}. Both plots show linear relation between $S$ index and $\lambda 8542$ indices, with $R^{+}$ is less scattered than $R$ index as the basal photospheric contribution was removed. 

\begin{figure}[!htbp]
    \centering
    \includegraphics[scale=0.7]{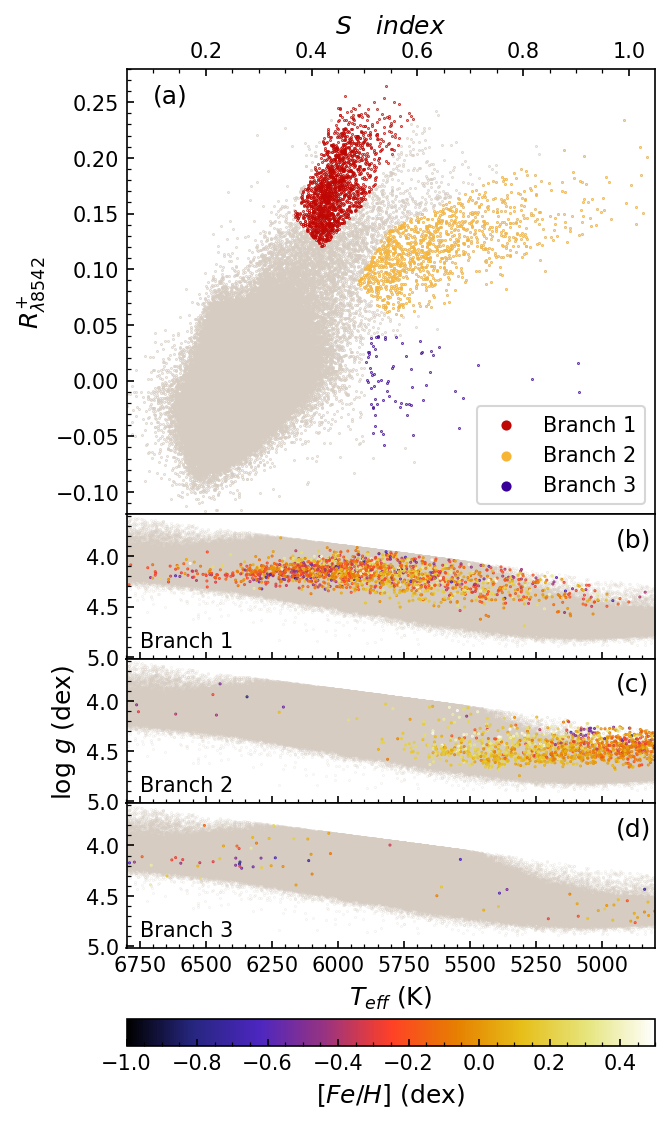}
    \caption{(a): Distribution of $S$ - $R^{+}_{\lambda 8542}$, different branches are defined by eyes and plotted in different color,as indicated in the frame. The background stars are shown in grey. (b): Distribution of Branch 1 stars in stellar parameter space. Metallicity is indicated by color, as shown in the bottom color bar, (c): Distribution of Branch 2 stars. (d): Distribution of Branch 3 stars.}
    \label{fig10}
\end{figure}

Visually inspecting Figure \ref{fig9}, the high activity index star seems to be divided into 3 branches. We label
those 3 branches in Figure \ref{fig10} and plot their distributions in stellar parameter space in the lower panels of Figure \ref{fig10} respectively. 
For Branch 1, we did not find any specific tendencies in the distribution of $T_{eff}$, $[Fe/H]$, but they almost located at $\text{log} \ g < 4.5$. Branch 2 has lower $R^{+ }$ index than Branch 1 and extend to very high $S$ index end. They distributed at temperatures below 5750K and exhibits high metallicity. Branch 3 has high $S$ index but low $R^{+ }_{\lambda 8542}$ index.
The sample size of Branch 3 is small, but they has a broad temperature range. They shows high metallicity in the low-temperature end and low metallicity in the high temperature end.


\begin{figure}[!htbp]
    \centering
    \includegraphics[scale=0.33]{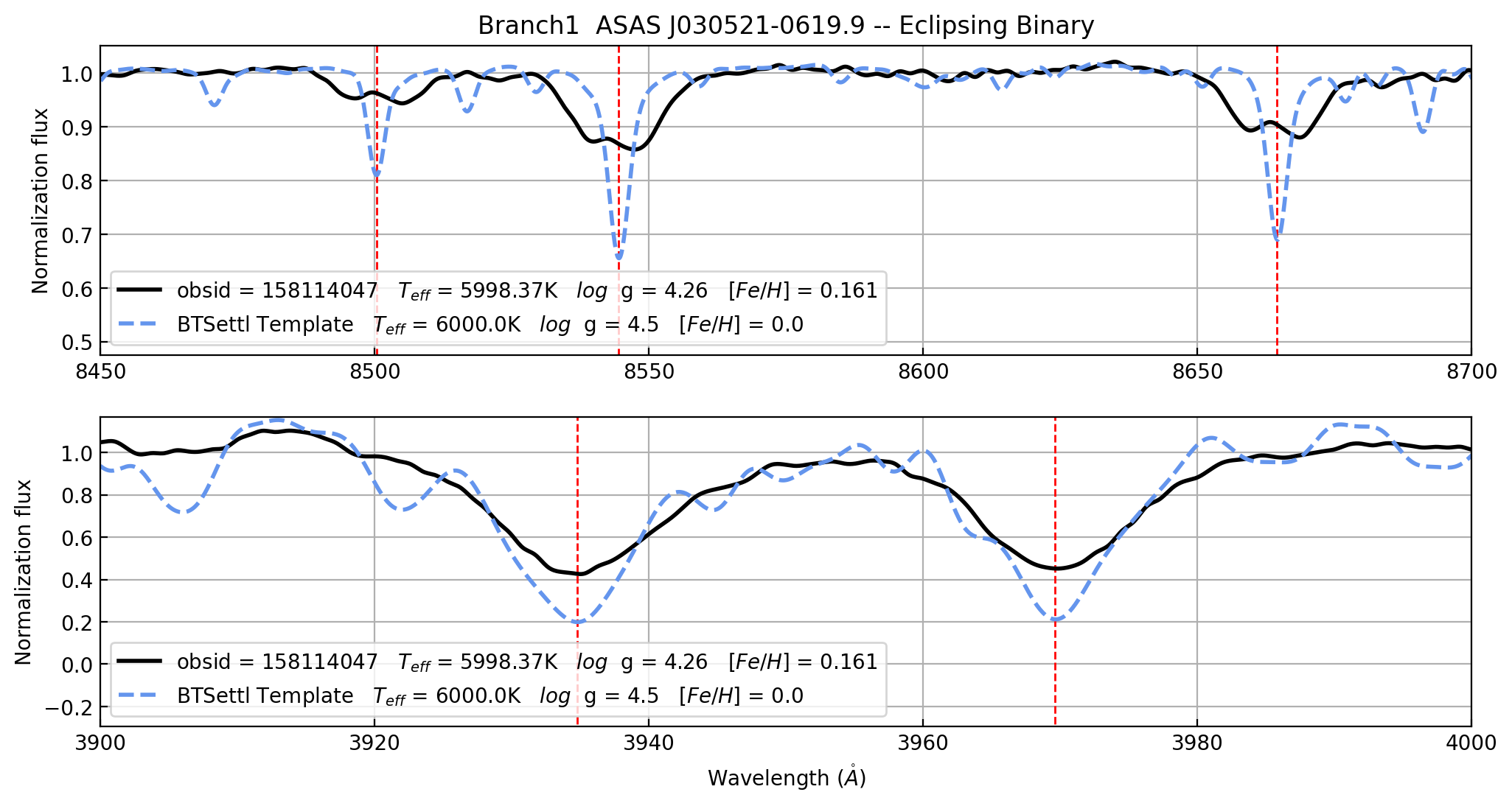}
    \includegraphics[scale=0.33]{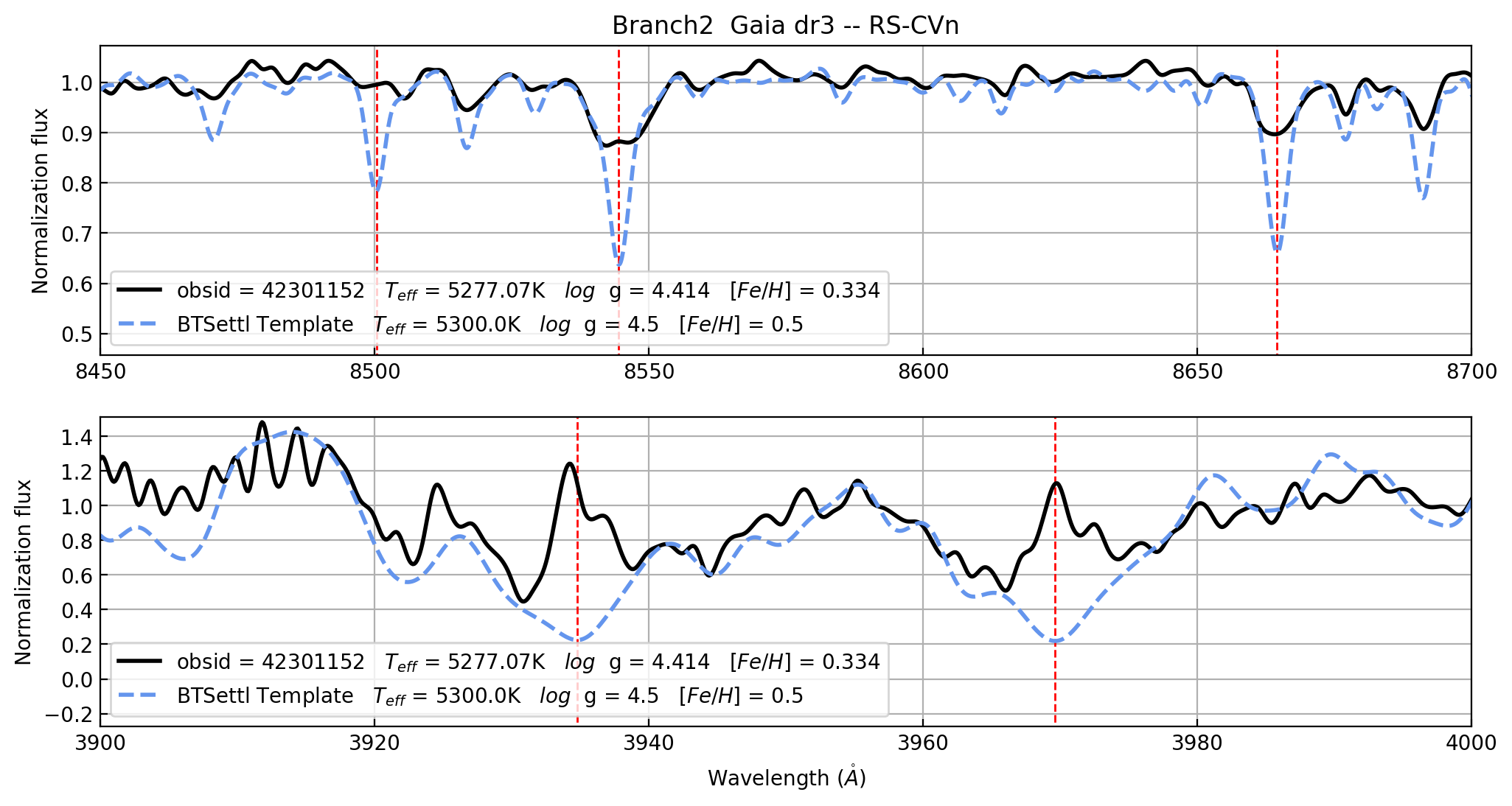}
    \includegraphics[scale=0.33]{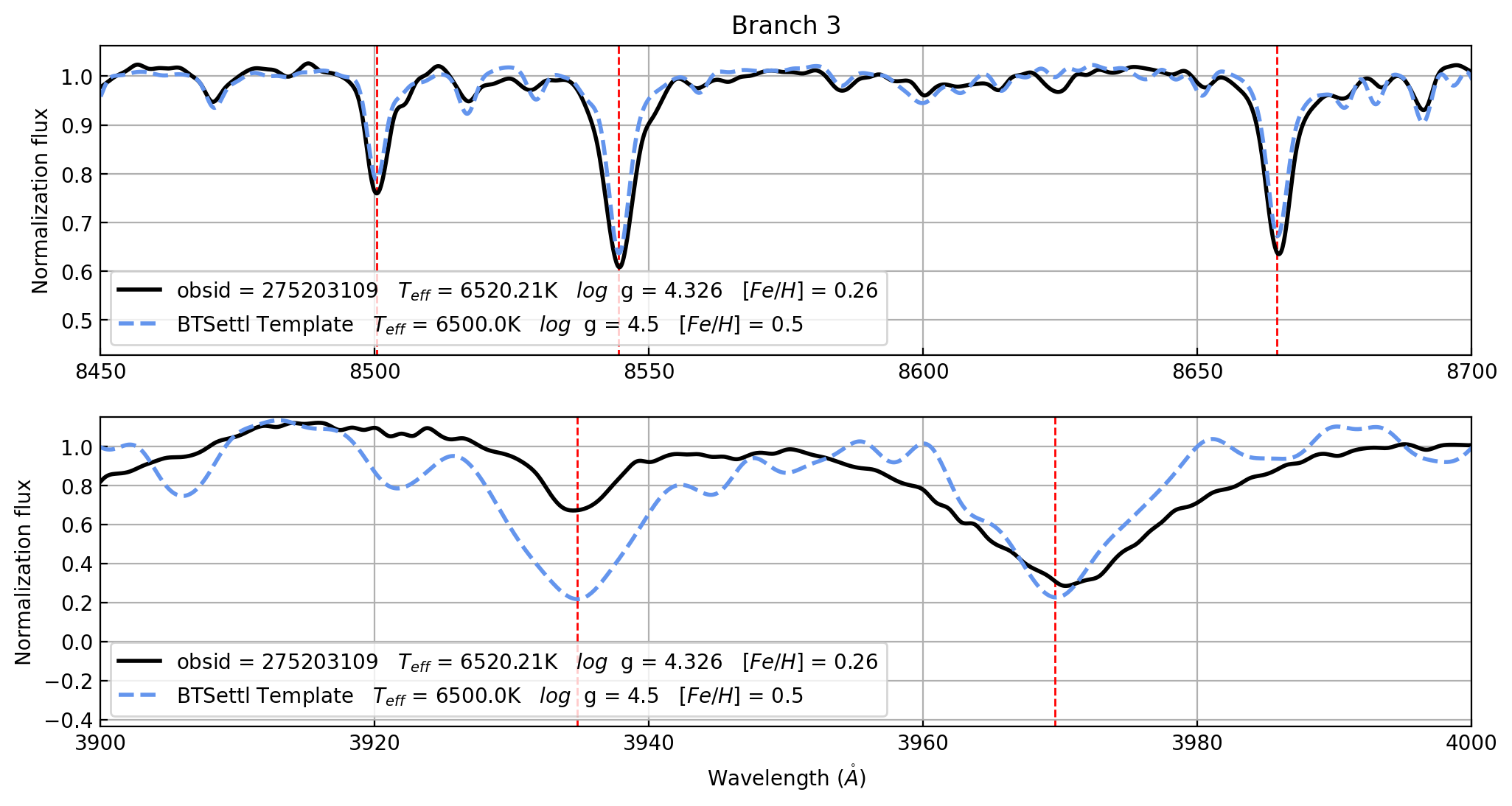}
     \caption{The typical spectra from the 3 branch star in Figure \ref{fig10}. From top to bottom are a star of Branch 1,2 and 3 respectively. The spectra and the corresponding template of the Ca \footnotesize{II} IRT region and the Ca {\footnotesize{II}} H\&K region are plotted respectively.}
    \label{3brach}
\end{figure}

To investigate the properties of the 3 groups, we check the spectra by eyes, the typical spectra are show in Figure \ref{3brach}.

\begin{figure}[!htbp]
    \centering
    \includegraphics[scale=0.65]{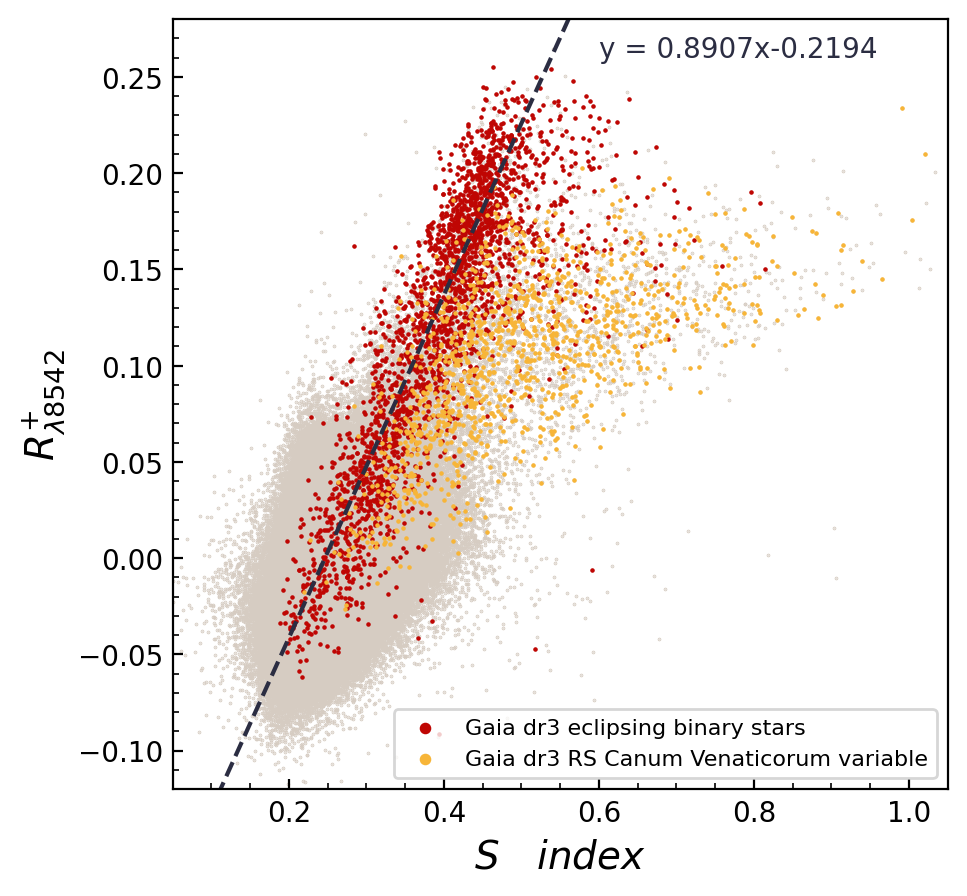}
    \caption{ $R^{+}_{\lambda8542}$ VS $S$ index distribution. The background grey points are the same as Figure \ref{fig9}. Overlaid color samples
     are stars cross matched with gaia dr3 eclipsing binaries and RS-CVn stars, respectively, as indicated in the figure. Those two samples are coincident with Brach 1 and Brach 2 defined in Figure \ref{fig10}, respectively.
    The black dash line is the linear regression of the eclipsing binary sample, the function is marked in the upper right corner.}
    \label{fig12}
\end{figure}

\begin{figure*}[!htbp]
    \centering
    \includegraphics[scale=0.78]{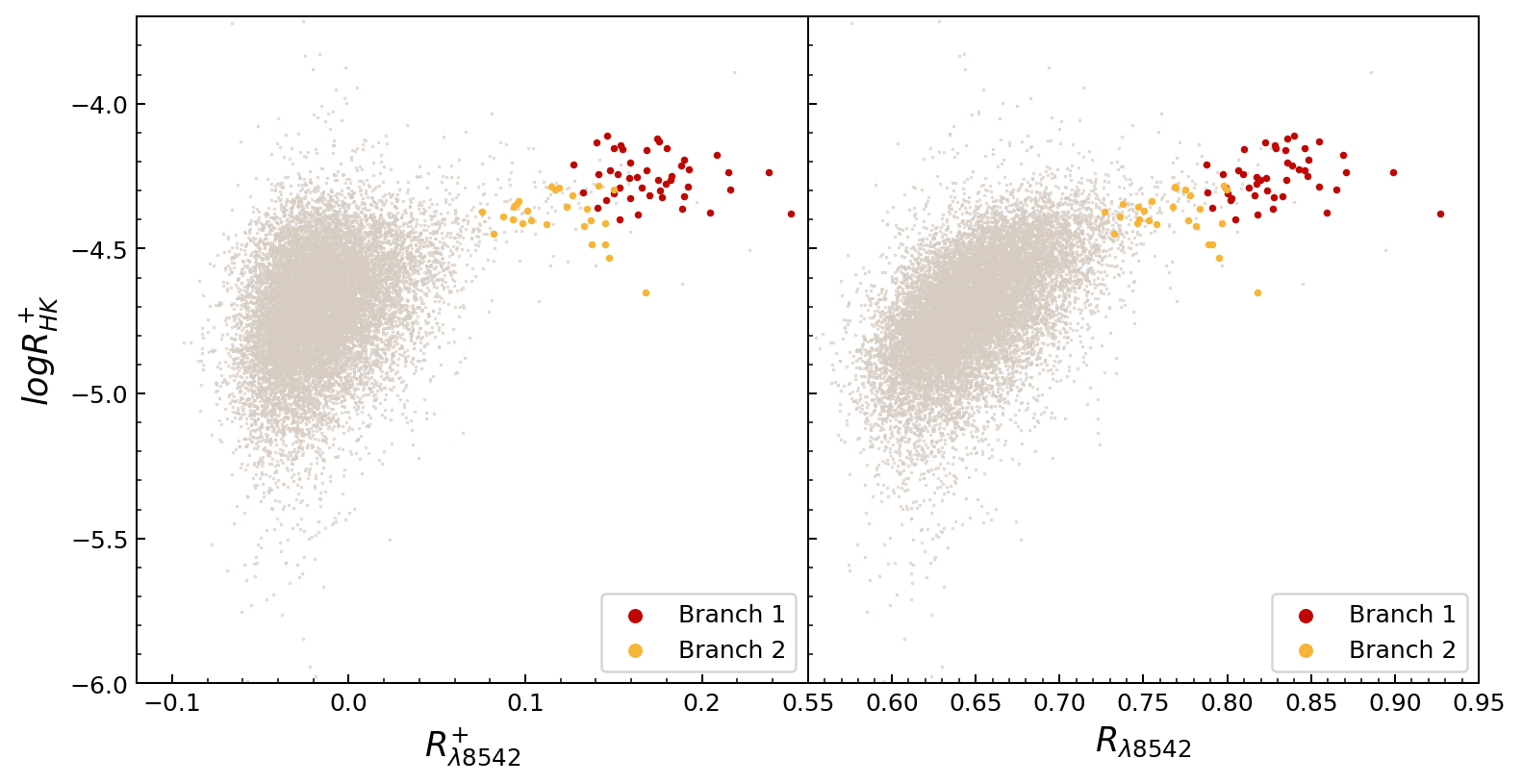}
    \caption{left panel: Relationship of $R^{+}_{\lambda 8542}$ - $log R^{+}_{HK}$. right panel: Relationship of $R_{\lambda 8542}$ - $log R^{+}_{HK}$. Red and yellow points are stars of Branch 1 and 2 defined in \ref{fig10}, respectively.}
    \label{fig11}
\end{figure*}

\begin{enumerate}

    \item Most of the spectra in Branch 1 show character of double lines at IRT band and H\&K lines are broader than the template, which is typical in spectral binaries. For LAMOST low resolution spectra, the radial velocity separation should be more than 150km/s for the 2 lines to be clearly  discerned. So those are highly possible to be close binaries with larger RV difference and similar luminosity.
    To confirm this conclusion, we cross-match the samples with the \texttt{gaiadr3.vari\_eclipsing\_binary} catalog \citep{gaia2016gaia, vallenari2023gaia, mowlavi2023gaia}, which yielded 1727 common spectra (1507 common sources).
    About 66\% (997/1507) stars in our defined Branch 1 region(Figure \ref{fig10}) are coincident with the 
    Gaia dr3 eclipsing binaries. As the Gaia samples are selected by light curves thus are highly dependent on the inclination angle, the rest 34\% of Branch 1 are either spectral binaries of low inclination that show no eclipse in lightcurve, or possibly some single stars show real high activity. So a large fraction of this branch should be close binaries mimic the chromospheric emission due to the index calculation algorithm, most of them are not active stars, or at least not as high as the $S$ or $R$ index indicated. Further investigation is necessary to determine their nature.
    The Gaia eclipsing binary sampleS extend linearly to the low active index end 
    in the $S$ vs $R^{+}$ plot (Figure \ref{fig12}). We fitted the Gaia samples with RANSAC (Random Sample Consensus) regression algorithm provided by \texttt{sklearn} package \citep{scikit-learn}, the result is shown in Figure \ref{fig12}. 
    
    
    \item For Branch 2, we observed obvious emission cores in most of the spectra at the $H\&K$ lines, and filled-in core of the Ca {\footnotesize{II}} IRT absorption lines. So the branch 2 is dominated by highly chromospheric activity stars. From the parameters distribution, they are predominantly metal rich (i.e. $[Fe/H] > -0.2$) cool stars ($T_{eff} < 5700K$).
     As RS Canum Venaticorum variable (RS-CVn) is a type of chromospheric active binaries, we crossed correlate our catalog with the {\it Gaia} RS-CVn catalog \citep{rimoldini2023gaia}, and got 1187 spectra (1037 stars) in common. The matched Gaia RS-CVns are plotted in Figure \ref{fig12}, most of these stars are consistent with Branch 2 and exhibit clear differences from eclipsing binaries. 
     
    \item Branch 3 shows higher $S$ index and relatively lower $R^{+}$ index, which means the H\&K lines show higher activity than IRT lines. We suspect this may be caused by binaries or visually close stars with different temperature, the blue and red region are dominated by different star that falling into the same fiber. Since there are only 70 stars in this category, those stars were checked one by one. Miscellaneous informations such as CDS (\url{http://cdsportal.u-strasbg.fr/}) images, SED, LAMOST spectra, \texttt{Gaia non-single star} list \citep{2023A&A...674A..10H}, TESS light curves \citep{ricker2015transiting, sullivan2015transiting} and Kepler light curves \citep{howell2014k2} were collected to help to understand the nature of these targets, those information are listed in the last column of Table \ref{tabA1}. From the table, 21 of them are binaries or spacial coincidence, supporting our speculation. 27 of them are variables that may show high activity, 10 of them have no particular reason and the rest of them are due to pollution or are with wrong spectral type, further study is necessary to know their nature.    
    
    
\end{enumerate}

\subsection{Comparing with \texorpdfstring{$R^{+}_{HK}$} \ index}

Using the $log R^{+}_{HK}$ index in \citet{zhang2020magnetic} (Table \ref{tab1}), the distribution of $R^{+}_{\lambda 8542}$ - $log R^{+}_{HK}$ and $R_{\lambda 8542}$ - $log R^{+}_{HK}$
were plotted in Figure \ref{fig11}. The stars in Branch 1
and Branch 2 in previous section are also plotted. 
The overall distribution is
similar to \citet[see][Fig.7]{vzerjal2013chromospherically}, though they used $EW_{IRT}$ as infrared activity indicator.
Binary stars are more likely to appear in the region of both high H\&K and high IRT index, similar to \citet{vzerjal2013chromospherically}.
From Fig \ref{fig12} and \ref{fig11}, high H\&K or high IRT index alone is not a good indicator of activity 
as there is a large population of binaries with low stellar activity mimic the high activity star. Combing the H\&K index and the IRT index, especially in the S vs $R^{+}_{\lambda 8542}$ distribution plot, will be more helpful to decern different population of stellar activity.


\section{Conclusion}
We defined new infrared Ca {\footnotesize{II}} triplet stellar activity indices, $R$ and $R^{+}$, and derived the indices for 699,348 spectra of 562,863 solar like F, G and K-type stars.
These activity indices, as well as their estimated uncertainties and other basic information are integrated in a database available at \url{https://nadc.china-vo.org/res/r101246/}. 

 Comparing the indices of $\lambda8498$, $\lambda8542$ and $\lambda8662$ lines, they show linear correlation in each pairs. The $ R^{+}_{\lambda8542}$ is the strongest among the three lines, and could be used as the indicator to represent the Ca {\footnotesize{II}} IRT activity. 
 We presented the distribution of $\lambda8542$ index in stellar parameter space, and selected samples of high active and low active stars, respectively. The fraction of low active stars decrease with the temperature, well the fraction of high active stars first decrease with the temperature above 5800K, then below 5800K, the fraction increase with the decreasing temperature. 
 We further compare our infrared activity index with the Ca {\footnotesize{II}} H\&K index and find that the high $S$ index star could be divide into 3 branches, Branch 1 are mostly spectral binaries with double lines that mimic the emission line core, Branch 2 are RS CVns that show high activity, Brach 3 are stars show high $S$ index but relatively low $R^{+}_{\lambda8542}$ index due to difference reasons.
 Combining the Ca{\footnotesize{II}} H\&K $S$ index and $R^{+}_{\lambda 8542}$ is particularly useful in selecting true chromospheric active stars. A future work is necessary to exclude pollution from low active binaries and establish a pure sample of high active stars.
 



\section*{ACKNOWLEDGEMENTS}

This work is supported by the National Key R\&D Program of China (2019YFA0405000).
 Z.H.T. also thanks the support
of the National Key R\&D Program of China 2022YFA1603002, NSFC 12090041, NSFC 11933004 and NSFC 12273056.  

This work has made use of data from the Guoshoujing Telescope(the Large Sky Area Multi-Object Fiber Spectroscopic Telescope, LAMOST). LAMOST is operated and managed by the National Astronomical Observatories, Chinese Academy of Sciences. (\url{http://www.lamost.org/public/?locale=en}). Funding for the LAMOST has been provided by the National Development and Reform Commission.

This work has made use of data from the European Space Agency (ESA) mission
{\it Gaia} (\url{https://www.cosmos.esa.int/gaia}), processed by the {\it Gaia}
Data Processing and Analysis Consortium (DPAC,
\url{https://www.cosmos.esa.int/web/gaia/dpac/consortium}). Funding for the DPAC
has been provided by national institutions, in particular the institutions
participating in the {\it Gaia} Multilateral Agreement. 

This research has made use of the SIMBAD database, operated at CDS, Strasbourg, France.

This paper includes data collected by the Kepler mission and obtained from the MAST data archive at the Space Telescope Science Institute (STScI). Funding for the Kepler mission is provided by the NASA Science Mission Directorate. STScI is operated by the Association of Universities for Research in Astronomy, Inc., under NASA contract NAS 5–26555.

This paper includes data collected with the TESS mission, obtained from the MAST data archive at the Space Telescope Science Institute (STScI). Funding for the TESS mission is provided by the NASA Explorer Program. STScI is operated by the Association of Universities for Research in Astronomy, Inc., under NASA contract NAS 5–26555.

{\it Facilities}: LAMOST, GAIA, TESS, Kepler

{\it Software}: Astropy \citep{robitaille2013astropy, price2018astropy, price2022astropy}, Astroquery \citep{2019AJ....157...98G}, SciPy \citep{2020SciPy-NMeth}, NumPy \citep{harris2020array}, Scikit-learn \citep{scikit-learn}, Matplolib \citep{Hunter:2007}, TOPCAT \citep{2005ASPC..347...29T}, Lightkurve \citep{2018ascl.soft12013L, brasseur2019astrocut}

\bibliography{refpaper}
\bibliographystyle{aasjournal}




\appendix

\renewcommand\thetable{\Alph{section}\arabic{table}}   
\section{List of Branch 3 stars}
\setcounter{table}{0}

\begin{longtable*}{|l|l|l|l|l|l|l|l|l|}

    \caption{Information of stars in Branch3 \label{tabA1}} \\
    
    \hline
    \textbf{No.} &\textbf{obsid} & \textbf{gaia\_source\_id} & \textbf{g\_mag} & \textbf{ra\_obs} & \textbf{dec\_obs} & \textbf{$R^{+}_{8542}$} & \textbf{S Index} & \textbf{Class}                     \endfirsthead 
    \hhline{|=========|}
    1& 181415234      & 2742433723412879360       & 13.07488                    & 1.883999         & 5.700471          & 0.023102               & 0.62965          & *UV excess/binary?                                  \\ 
    \hline
    2& 255415044      & 390549008386598016        & 14.26488                    & 11.21287         & 48.28147          & 0.035161               & 0.604901         & *Bright Star Pollution             \\ 
    \hline
    3& 182715182      & 376725089206420864        & 14.22984                    & 16.69563         & 44.43003          & 0.01305                & 0.554792         & Variable (G)  \\ 
    \hline
    4 & 209103032      & 114150201980200960        & 12.09423                   & 40.99757         & 24.91994          & -0.02086               & 0.585086         & *Nearby Star Pollution     \\ 
    \hline
    5& 162403203      & 108894639478505472        & 13.62411                    & 46.7181          & 22.2539           & 0.015495               & 0.903646         & *Young Stellar Object Candidate   \\ 
    \hline
    6& 157302145      & 125962495916228992        & 12.85282                    & 50.73753         & 34.37201          & -0.00395               & 0.61698          & Variable(G)                            \\ 
    \hline
    7& 286103110      & 67691055407537792         & 13.16509                    & 53.43127         & 23.15588          & -0.01764               & 0.610353         & Binary (G)           \\ 
    \hline
    8& 307915107      & 3250965204243797760       & 12.7828                     & 55.21712         & -1.54672          & -0.01576               & 0.685768         & *Visual Binary                     \\ 
    \hline
    9 & 111607167    & 70286319462343808          & 11.74727                    & 56.30801         & 26.5884           & 0.038976               & 0.525731         & Variable(G)                          \\ 
    \hline
    10 & 480603181      & 65205166993246080         & 14.21742                    & 56.58174         & 23.91762          & -0.04815               & 0.534726         & *Bright Star Pollution                    \\ 
    \hline
    11 & 100904105      & 65223618172733952         & 11.95708                    & 56.6641          & 24.02969          & 0.037559               & 0.539393         & *BY Dra Variable                   \\ 
    \hline
    12 & 204105048      & 163600634362268800        & 11.44613                    & 60.13293         & 27.42786          & 0.031089               & 0.515465         & *MS+WD Binary \footnote[1]{\citet{ren2020white}}                      \\ 
    \hline
    13 & 273916194      & 232362820257069440        & 10.37425                    & 62.41018         & 43.59254          & -0.0312                & 0.581389         & Variable(K)                          \\ 
    \hline
    14 & 470205184      & 232914736434443136        & 14.58639                    & 64.16136         & 45.60959          & -0.05821               & 0.536967         &                                    \\ 
    \hline
    15 & 384509039      & 3285027799594151680       & 13.08165                    & 65.08137         & 5.838964          & 0.039738               & 0.526472         & Variable(K)                           \\ 
    \hline
    16 & 275203109      & 253742995657660288        & 10.77930                    & 67.0192          & 45.56416          & -0.04711               & 0.517125         & Binary (G)                            \\ 
    \hline
    17 & 361716215      & 277067485569047680        & 11.67850                    & 67.28308         & 55.21747          & -0.00102               & 0.505846         & Variable(K)                        \\ 
    \hline
    18 & 250801006      & 3405685422487373568       & 14.17399                    & 73.94313         & 17.28189          & 0.039493               & 0.537972         & Variable (G)             \\ 
    \hline
    19 & 39104099       & 205354966385794048        & 12.32107                    & 73.95885         & 43.69652          & -0.04074               & 0.520087         & Variable(K)                          \\ 
    \hline
    20 & 528007141      & 3228908790535918976       & 14.17989                    & 75.26918         & 1.364525          & 0.000329               & 0.53148          & Variable(G)  \\ 
    \hline
    21 & 307304141      & 211681178338056192        & 12.04812                       & 78.64506         & 45.42125          & -0.04892               & 0.592449      & Variable(K)                            \\ 
    \hline
    22 & 678513097      & 281149010170791552        & 14.78009                    & 79.71296         & 59.046            & 0.005221               & 0.513852         & Variable(G)              \\ 
    \hline
    23 & 89713095       & 3448967285402131712       & 12.44196                    & 82.39802         & 32.74561          & 0.028367               & 0.520001         & *Nearby Star Pollution              \\ 
    \hline
    24 & 208806168      & 3333163830247192064       & 12.34049                    & 84.0461          & 6.520935          & 0.005564               & 0.517577         & Variable(K)                          \\ 
    \hline
    25 & 393309119      & 3397615659976935296       & 11.66793                    & 84.67204         & 18.00152          & -0.02442               & 0.516078         &                                    \\ 
    \hline
    26 & 505215137      & 3216524342533541248       & 15.34401                    & 85.14597         & -2.19503          & 0.013964               & 0.715127         & *Nebula Pollution               \\ 
    \hline
    27 & 505204206      & 3216417655546088192       & 15.33606                    & 85.17643         & -2.84414          & -0.04222               & 0.678071         & *Nebula Pollution                  \\ 
    \hline
    28 & 297011180      & 189407787175600640        & 11.11371                    & 85.21744         & 37.46183          & -0.00455               & 0.518133         & Variable(K)                          \\ 
    \hline
    29 & 505215105      & 3216425077249628544       & 14.59378                    & 85.24724         & -2.74513          & -0.02399               & 0.567592         & *Nebula Pollution                \\ 
    \hline
    30 & 547505087      & 3399231147498442112       & 11.37671                    & 85.84326         & 19.40144          & -0.00609               & 0.590741         & *Eclipsing Binary                  \\ 
    \hline
    31 & 127806031      & 3431388156057648128       & 11.96078                    & 90.89316         & 28.81194          & -0.02395               & 0.516935         & *Chemically Peculiar Star/Nearby star Pollution            \\ 
    \hline
    32 & 486302167      & 3423621618233438080       & 14.50957                    & 90.98312         & 21.90916          & -0.03055               & 0.52059          &                                    \\ 
    \hline
    33 & 606202112      & 3375271831352365568       & 13.13685                    & 91.74336         & 21.03054          & 0.033773               & 0.512433         & Variable(K)                                   \\ 
    \hline
    34 & 501916045      & 3328461188953301120       & 14.17653                    & 91.90313         & 8.044743          & 0.011596               & 0.510486         &                                    \\ 
    \hline
    35 & 641111226      & 3345719467060096768       & 15.06539                    & 92.92255         & 15.19583          & 0.024583               & 0.606389         &                                    \\ 
    \hline
    36 & 606211049      & 3426827038226866176       & 13.94504                    & 93.65705         & 25.50473          & -0.00978               & 0.542015         & *RR Lyrae Variable                 \\ 
    \hline
    37 & 486809137      & 3425551463001195008       & 11.62234                    & 94.2737          & 23.42387          & 0.02181                & 0.579099         & Variable(K)                          \\ 
    \hline
    38 & 267811169      & 3370935975970328192       & 14.46051                    & 95.64267         & 19.2116           & -0.0069                & 0.512879         &                                    \\ 
    \hline
    39 & 696613240      & 3102733650797714816       & 12.25146                    & 103.3226         & -3.61788          & -0.045                 & 0.531789         & Giants with wrong logg (L)            \\ 
    \hline
    40 & 378105061      & 993779054893891840        & 12.59996                    & 104.1611         & 54.1417           & 0.038976               & 0.560594         & Variable(K)                        \\ 
    \hline
    41 & 88605186       & 3109933798391183232       & 12.26119                    & 110.448          & -1.2325           & 0.039448               & 0.521697         & Variable(K)                   \\ 
    \hline
    42 & 88805176       & 3109936826350414592       & 10.76796                    & 110.5126         & -1.13856          & -0.02079               & 0.512853         & *Spectral Binary                \\ 
    \hline
    43 & 226703189      & 892715622559710592        & 14.19901                    & 113.7437         & 33.00618          & 0.016696               & 0.503644         & *MS+WD Binary \footnote[2]{\citet{gentile2015independent}}                       \\ 
    \hline
    44 & 93609075       & 3064639245085801344       & 12.01463                    & 123.5568         & -5.45447          & -0.00035               & 0.525129         & *Visual Binary/Variable(K)        \\ 
    \hline
    45 & 308415140      & 3098139547613310720       & 12.97677                    & 124.1939         & 8.390326          & 0.033797               & 0.570646         & Variable(K)                          \\ 
    \hline
    46 & 656613008      & 636182586087691392        & 14.07274                    & 136.9056         & 18.83429          & -0.01988               & 0.569886         &                                    \\ 
    \hline
    47 & 201907064      & 3824325436834913920       & 11.80175                    & 146.0101         & -4.58495          & 0.025776               & 0.508454         & Variable(G)                         \\ 
    \hline
    48 & 303015088      & 830588577026980992        & 11.67706                    & 160.1215         & 46.73302          & 0.011674               & 0.552609         & *High Proper Motion Star          \\ 
    \hline
    49 & 401214096      & 3816910296057695232       & 12.01111                    & 168.8487         & 5.573148          & 0.021652               & 0.521994         &  SB1 (L)                              \\ 
    \hline
    50 & 208509165      & 3695446967363569408       & 12.37765                    & 188.6458         & -1.01727          & 0.007846               & 0.51159          &  *Visual Binary                      \\ 
    \hline
    51 & 132212074      & 3650688086675908352       & 12.20117                    & 221.798          & -0.49315          & 0.029954               & 0.641464         & *Hot Subdwarf             \\ 
    \hline
    52 & 426805127      & 1597737184257054720       & 12.58121                    & 233.8434         & 53.58372          & 0.001588               & 0.81768          & Variable(K)                          \\ 
    \hline
    53 & 152601123      & 1353107529388288896       & 13.30084                    & 252.5303         & 40.17427          & -0.04238               & 1.122254         & Cosmic Ray Pollution(L)              \\ 
    \hline
    54 & 334701053      & 1360809745779585152       & 12.16614                    & 262.6004         & 44.48631          & 0.025941               & 0.5065           & Variable(K)                          \\ 
    \hline
    55 & 574714131      & 2133632795086109440       & 14.42111                    & 286.6875         & 50.6358           & -0.01803               & 0.558479         &               \\ 
    \hline
    56 & 243012154      & 2102151990479456128       & 12.88417                    & 287.9217         & 41.05147          & -0.00226               & 0.542695         & *Nearby Star pollution                         \\ 
    \hline
    57 & 369703082      & 2099502579773618560       & 12.64614                    & 289.1561         & 39.14371          & 0.000118               & 0.505672         & *Visual Binary                          \\ 
    \hline
    58 & 52403133       & 2101074331648268032       & 13.70564                    & 290.483          & 39.73531          & -0.01383               & 0.546077         &                                    \\ 
    \hline
    59 & 580505166      & 2052645379929910144       & 13.74717                    & 290.9112         & 38.33558          & 0.023596               & 0.509367         & Variable(G)                         \\ 
    \hline
    60 & 362811058      & 2134979074057185408       & 13.58594                    & 295.6626         & 50.14518          & 0.03832                & 0.512928         & *Rotating Variable/Visual binary                 \\ 
    \hline
    61 & 355104179      & 2079247720169124992       & 14.78923                    & 299.1187         & 45.4898           & 0.007736               & 0.623634         & *Pulsating Variable                 \\ 
    \hline
    62 & 158908013      & 2082103770340838144       & 13.23867                    & 300.7669         & 44.86653          & -0.02103               & 0.578636         & *Rotating Variable                \\ 
    \hline
    63 & 260702136      & 2068072279678698112       & 13.12818                    & 306.8594         & 41.61058          & 0.010496               & 0.588413         & *Nearby Star Pollution? Variable(G)                         \\ 
    \hline
    64 & 587915134      & 2163026176885565568       & 14.99146                    & 314.1843         & 44.80388          & -0.00931               & 0.670712         & *Nebula Pollution                  \\ 
    \hline
    65 & 169005207      & 1781458323057855360       & 12.24709                    & 331.3152         & 20.28735          & 0.026222               & 0.625867         & *Visual Binary                          \\ 
    \hline
    66 & 75308136       & 2735568299794226304       & 11.86591                    & 335.6045         & 15.65098          & -0.00043               & 0.544412         & *Visual Binary                     \\ 
    \hline
    67 & 75308129       & 2735577375059931264       & 12.50423                    & 335.722          & 15.78769          & -0.01789               & 0.535658         &                                    \\ 
    \hline
    68 & 270405145      & 2008973392251158784       & 14.50971                    & 345.2708         & 55.97517          & -0.02235               & 0.587842         & *Nearby Star Pollution             \\ 
    \hline
    69 & 387904014      & 2664836171318493696       & 14.05540                    & 349.4505         & 7.379211          & 0.025358               & 0.609652         & *Hot Subdwarf Candidate / UV excess     \\ 
    \hline
    70 & 180206182      & 1924190810839989632       & 12.92156                    & 350.0968         & 40.73101          & -0.0105                & 0.905189         & *Visual Binary                         \\
    \hline
    
\end{longtable*}

\tablecomments{In the classification, asterisks "*" indicate sources that have been identified through the visual examination of relevant information in CDS website, such as SIMBAD information, literatures, SEDs and images; "Varaiable (G)" and "Binary (G)" are targets cross matched with \texttt{gaiadr3 variability} \citep{rimoldini2023gaia} and \texttt{gaiadr3 non-single stars} \citep{2023A&A...674A..10H} databases respectively; "Varable(K)" stands for stars show apparent variations in Kepler or Tess light curves by visual inspection, as some of them showing periodic variations, binaries could not be excluded. "(L)" means the judgement is derived by inspecting the LAMOST spectra.} 

\end{document}